\documentclass[usenatbib]{mn2e}

\usepackage{graphicx,amsmath,amssymb,subfigure}

\def\simgt{\mathrel{\lower0.6ex\hbox{$\buildrel {\textstyle >}
 \over {\scriptstyle \sim}$}}}
\def\simlt{\mathrel{\lower0.6ex\hbox{$\buildrel {\textstyle <}
 \over {\scriptstyle \sim}$}}}

\newcommand{\mc}{\multicolumn}

\begin{document}

\title[The evolution of the radio source population]{The evolution of
radio sources in the UKIDSS-DXS XMM-LSS field}

\author[McAlpine \& Jarvis]
{Kim McAlpine$^{1}$ and Matt J. Jarvis$^{2,3}$\\
\footnotesize
$^{1}$Department of Physics and Electronics, Rhodes University, Grahamstown,
6139, South Africa\\
$^{2}$Centre for Astrophysics, Science \& Technology Research Institute,
University of Hertfordshire, Hatfield, Herts, AL10 9AB, UK\\
$^{3}$Physics Department, University of the Western Cape, Cape Town, 7535, South Africa\\
}

\maketitle

\begin{abstract}

We investigate the cosmic evolution of low luminosity ($L_{\rm{1.4GHz}}<10^{25}\rm{W~Hz^{-1}sr^{-1}}$) radio sources in the XMM
 Large Scale Structure survey field (XMM-LSS). We match low frequency selected (610~MHz) radio sources in the XMM-LSS field with near infrared $K$-band observations over the same field from the UKIRT Infrared Deep Sky Survey. 
We use  both the mean $V/V_{\rm{max}}$ statistic and the radio luminosity function  of these matched sources to quantify the evolution of the
 co-moving space density of the 
low luminosity radio sources in our sample. Our results indicate that the low luminosity sources evolve differently to their
 high luminosity counterparts out to a redshift of z$\sim$0.8. The derived luminosity function is consistent with an increase in the co-moving space density
of low luminosity sources by a factor of $\sim$1.5 at z=0.8. We show that the use of the $K-z$ diagram for the radio source population, although coarser than a full photometric redshift analysis, produces consistent results with previous studies using $\sim >10$ band photometry. This offers a promising method for conducting similar analyses over the whole sky with future near- and mid-infrared surveys.
\end{abstract}


\section{Introduction}\label{sec:intro}
The strong cosmic evolution of the most powerful radio sources was first deduced by \citet{Longair1966} from low frequency radio source counts.
This evolution has  since been confirmed by a number of investigations which imply that the co-moving space density of high luminosity radio sources has decreased by 
a factor of approximately 1000 between  $z\sim2$ and $z\sim0$ (e.g. \citealp*{LRL1983}; \citealp{DP1990,Willott2001}).  Beyond $z\sim 2$ the evolution remains uncertain \citep[e.g.][]{JarvisRawlings00,Jarvis01b} but appears to undergo a gradual decline \citep[e.g.][]{Wall05}.
The evolution of the low-power radio sources is less well constrained, but early studies of the radio source counts indicated that 
the low luminosity sources could not be evolving as strongly as their high luminosity counterparts (\citealt{Longair1966}; \citealt*{DLZ1970}). Many models of the evolution of radio sources 
thus divide the radio source population into two independently evolving components,
a strongly evolving high luminosity component and a lower luminosity component with much weaker evolution (e.g. \citealt{JW1999,Willott2001}).

Radio galaxies can be classified into two morphologically distinct groups identified as Fanaroff-Riley class I (FRI)
 and  Fanaroff-Riley class II (FRII) sources \citep{FR1974}.  The FRI sources have  higher surface brightness close to the centre of their radio lobes whereas the
 FRII sources have  highly collimated large-scale jets and bright emission hotspots at the edges of their radio lobes. The FRII sources are typically more
luminous than the FRI sources with the division in luminosity falling at roughly $L_{\rm{1.4GHz}}=10^{25}~\rm{W~Hz^{-1}sr^{-1}}$. This dividing luminosity may be
dependent on the optical luminosity of the host galaxy, a higher optical host luminosity results in a higher FRI/FRII division luminosity \citep{LO1996}.
 It has been suggested that the FRI and FRII sources might respectively be associated with the slowly and rapidly evolving components of the 
radio source population \citep{JW1999}. However \citet{Rigby} find evidence that high luminosity FRI sources evolve as rapidly as FRII sources
of comparable radio power. Furthermore, \citet*{Gendre2010} performed a detailed comparison of the radio luminosity function of FRI and FRII  sources  
at a number of redshifts. This analysis revealed that both populations experience luminosity dependent number density enhancements at 
higher redshifts, z$\approx$0.8$\sim$2.5, and that 
there were no significant differences between the enhancements measured for these two populations.   
  These results would seem to suggest that both types of FR sources experience
 a common evolutionary history and thus
 cannot fully account for the observed dichotomy in cosmic evolution.

Recent evidence suggests that a more important division in the  radio source population, distinct from the morphological classification, may
be related to  different modes of  black hole accretion.   Radio galaxies have also been classified according to the presence
or absence of narrow high-excitation  emission lines in the spectra of their optical host galaxies \citep{HL1979,Laing1994,JR1997,Willott2001}. Objects without high-excitation
emission lines are referred to as low-excitation radio galaxies (LERG). Low luminosity FRI galaxies are predominantly LERG's and most objects with 
high excitation emission lines are associated with the more powerful FRII sources. However the relationship between FR class and the emission line classification
scheme is not one-to-one as many FRII galaxies have been found to be low-excitation radio galaxies \citep[e.g.][]{Evans2006}.  It has been argued that these two groups correspond to different AGN phenomena powered by fundamentally different
modes of accretion. The LERG sources are powered by a radiatively inefficient accretion of the hot gas in the intergalactic medium, referred to as ``radio'' mode accretion. While
the HERG are
powered by radiatively efficient accretion of cold gas, referred to as quasar mode accretion (\citealt{Evans2006}; \citealt*{Hardcastle2006,Hardcastle2007}; \citealt{Herbert2010}).
 \citet{Smol2009} argue
that the observed bimodal evolution of radio sources may be caused by differences in the evolution of these black hole accretion modes.  

Understanding the relationships between different classes of radio galaxies, and their cosmic evolution, is important for understanding models of
galaxy formation and evolution. Semi-analytic models of hierarchical galaxy formation predict that galaxies in the local universe  should be more 
massive and have
higher rates of star formation than is observed \citep{Bower2006}. There is increasing observational evidence that radio activity from AGN may
 disrupt cooling flows in 
 luminous early-type galaxies thus slowing or preventing significant accretion of gas and further star formation \citep{Fabian,Birzan2004,RJ2004}.
 This negative AGN feedback has been successfully incorporated into models of galaxy formation which are then able to better reproduce several 
features 
of the observed universe including the exponential cut-off in the bright end of the galaxy luminosity function \citep{croton2006,Bower2006}.
\citet{Best2006}  implied that the low luminosity radio sources may be the predominant contributors to this negative feedback effect.

Current studies of the cosmic evolution of low power radio sources seem to imply that these sources are not evolving at all or 
experience only mild negative evolution. \citet{Clewley2004} found
that the low luminosity radio sources with $L_{\rm{325MHz}}<10^{25}\rm{~W~Hz^{-1}sr^{-1}}$ exhibit no evolution out to redshift $z\sim0.8$, \citet{Waddington}
also found no evidence of evolution in their study of 72 sources. 
\citet{Sadler} find
evidence of mild evolution out to $z\sim 0.7$ in their comparison of the radio luminosity function of sources in the 2dF-SDSS Luminous Red Galaxy (2SLAQ) and
QSO survey with that of sources in the 6dF Galaxy survey (6dFGS). This  was consistent with pure 
luminosity evolution of the form $(1+z)^{2.0\pm0.3}$ which ruled out the no evolution scenario
at the 6$\sigma$ level. \citet*{Donoso} also find evidence of mild positive evolution in the $z=0.1\sim0.55$ redshift range. Using radio sources in the VLA-COSMOS
survey \citet{Smol2009} find evidence that this mild evolution continues out to $z\sim 1.3$. All of these studies imply that the level
of evolution in the low power radio sources is significantly less than that taking place in the high luminosity sources.

In this study we use low frequency radio sources detected in the XMM-Large Scale Structure survey field to investigate the co-moving space density of low-luminosity radio
sources. The low-frequency selection is preferred over high-frequency (e.g. $\geq 1.4$~GHz), as it provides an orientation independent selection, as the low-frequency detects the optically-thin lobe emission, whereas high-frequency surveys contain a higher fraction of pole-on sources where the optically thick core dominates \citep[e.g.][]{JM02}. 

In section 2 and 3 we discuss the radio and optical observations used in our analysis. Section 4 outlines the method we used to cross match the
radio sources with their optical counterparts. In section 5 we describe the method used to obtain redshifts for our sample. In the section 6 and 7
we discuss the $V/V_{\rm{max}}$ statistic used to quantify the evolution of our sources. In section 8 we determine the radio luminosity function of the low luminosity
sources in our observed field and use it to investigate their evolution. We present our conclusions in section 9.   

 We assume the $H_{0}$=70~km.s$^{-1}$.Mpc$^{-1}$ and a $\Omega_{M}$=0.3 and $\Omega_{\Lambda}$=0.7 cosmology throughout this paper.

\section{Radio data at 610~MHz}\label{sec:survey}
We use the radio observations of the XMM-LSS field described in detail in \citet{tasse610}. These observations cover 13 degrees$^{2}$ centered at
$\alpha$(J2000)=2$^{h}$24$^{m}$00$^{s}$ and $\delta$(J2000)=-4$^{\circ}$09$\arcmin$47$\arcsec$. This field was observed for a total of 18~h with
Giant Metrewave Radio telescope (GMRT) in August 2004 at a frequency of 610~MHz. 
The observations have 
a resolution of $\sim$ 6.7$\arcsec$ and a $5\sigma$ rms sensitivity ranging from  1.5~mJy to 2.5~mJy
across the observed field. \citet{tasse610} detect 767 radio sources in this field and obtain estimates or upper limits for the spectral
indices of these sources by comparison with the NRAO VLA Sky Survey (NVSS).

\section{$K$-band Data}

\begin{figure}
\includegraphics[width=0.95\columnwidth]{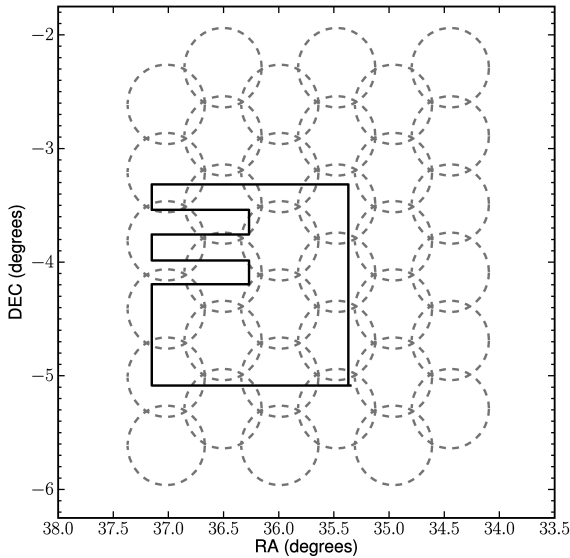}
\caption{
The location of the radio and infrared data used in this paper. The light grey circles show the 610~MHz pointings of the GMRT. The thick black line
indicates the UKIDSS DXS coverage of the XMM-LSS field in the 7th data release.}
\label{fig:cov}
\end{figure}

We match the radio sources detected in the XMM-LSS field with  $K$-band sources detected in the 7th data release (DR7) of the ongoing UKIRT
Infrared Deep Sky Survey (UKIDSS) \citep{Law}. UKIDSS uses the Wide Field Camera mounted
on the UK Infrared Telescope (UKIRT) \citep{Casali2007}. This survey, which began in May 2005,
comprises
5 complementary sub-surveys of varying depths and sky coverage. Our
study uses data from the Deep extragalactic survey (DXS) which is a deep, wide
survey 
over four observed fields. The survey is aiming to cover 35 degrees$^{2}$ to a planned
5$\sigma$ depth of K$\sim$20.8 \citep[see e.g.][]{Kim2010}.  We use the online catalogue of sources in the XMM-LSS field, the sky
coverage of this field in the 7th data release is illustrated in figure \ref{fig:cov},the overlap in the DXS sky coverage and the radio observations is
approximately 2.75 degrees$^2$. This
data is complete down to K$=$19.2 \citep{Kim2010} and varies in depth across the observed field. Thus to ensure completeness we adopt a cut-off of $K>19.2$ for the
 optical counterparts to our radio sources. K-band magnitudes in the UKIDSS survey are calibrated in the Vega system described by \citet{hewett}

\section{Radio and Infrared Matching}

For the purposes of identifying near-infrared counterparts to the radio data we divide the radio data into two categories.  
The first category comprises sources whose
 radio and optical emission are expected to be physically co-incident and includes all single component radio sources as well as  partially resolved 
multiple component sources and double radio
sources with clear radio cores. The second category contains  sources where the optical emission is expected to be separated from the
radio emission, this includes double radio sources with no detected radio core, radio jets and other complex multiple component structures. 

To determine the optimal pairing radius for sources in the first category we consider the accuracy of the radio and $K$-band source positions.
The radio source positions in the catalogue we use are accurate to approximately 2$\arcsec$ \citep{tasse610} and the
UKIRT survey provides accurate astrometry to within 0.1$\arcsec$ \citep{Law}.  We thus  adopt a pairing radius of  5$\arcsec$, corresponding to roughly
 2.5$\sigma$, where $\sigma$ is the combined error in the positions  of the radio and $K$-band sources. This large radius ensures relatively few
true optical counterparts will be located outside our search area.  However as the surface density of optical sources increases there is an increasing
probability that an an optical counterpart may appear within our adopted search radius by chance. 
To determine the probability of a given optical counterpart being such a  spurious alignment we use
the method of \citet{Downes1986}. This method  calculates the probability P  that an optical source
could lie at the given distance from the radio source by correcting the raw possionian probability 
 for the number of ways that such an apparently significant association could take place. We reject all matches whose corrected probability value
P$>$~0.05. As a further consistency check we plotted the contours of all the radio sources in DXS XMM-LSS field over the optical images 
and inspected the data by eye.

Determining the optical counterparts of the more complex radio sources in our second category is much less suited to statistical methods. Thus we
determined the best match for sources in this category by means of visual inspection. We find matches brighter than $K$=19.2 for 131 sources out of a
 total of 213 radio sources detected in the DXS XMM-LSS field.  

\section{Estimating redshifts}\label{reds}

We use a method of redshift estimation, developed by \citet{Cruz2007}, which utilises the tight correlation between $K$-band magnitude and
 redshift for radio galaxies. This K-z relation has been investigated for a number of radio galaxy samples including the
 3CRR \citep{lilly}, 6CE \citep{Eales1997}, 6C* \citep{Jarvis2001} and 7CRS samples \citep{Willott2003} and continues to at least z=3 \citep{Jarvis2001}.

The method developed in \citet{Cruz2007} uses a model of the  distribution of radio galaxies as a function of
redshift  together with the linear $K-z$ relation to generate a
 Monte Carlo simulation of radio sources which populate a synthetic $K-z$
diagram.  The advantage of this method over simply applying a linear fit to the $K-z$ relation is that it incorporates information about the scatter in
this relation and is thus able to provide a measure of the uncertainty in 
the output redshift estimates.

The distribution of radio galaxies  and $K-z$ relation used in our simulations were based on the combined dataset of the 3CRR,6CE,6C* and 7CRS samples.
\citet{Cruz2007} find the radio galaxies in this sample to be well fitted by a function of the form : \begin{equation}
                 n(z)=A\exp\{-[\Sigma_{i=0}^{5}a_{i}(\log_{10}{z})^{i} ]^{2}\}
                                                                                                                           \end{equation}

with polynomial coefficients : 
\begin{center}
\begin{tabular}{llll}
A=197.96; & $a_{0}$=-0.39; & $a_{2}$=1.00; & $a_{4}$=1.47;\\
× & $a_{1}$=1.17; & $a_{3}$=1.83; & $a_{5}$=0.38.
\end{tabular}
\end{center}

\citet{Willott2003} quote the  best fit to the $K-z$ relation in this combined sample as the second-order polynomial \begin{equation}K(z) = 17.37+4.53\log_{10}{z}+0.31(\log_{10}{z})^2.\end{equation}

The Monte Carlo simulations assume Gaussian deviations about a mean $K$-magnitude 
obtained from the $K-z$ relation. A constant dispersion for these deviations is adopted for the entire redshift range. We use the dispersion $\sigma_{k}$
obtained by \citet{Cruz2007} via a fit to the same dataset used to obtain the $K-z$ relation. This dispersion was found to be $\sigma_k$=0.593. 

\begin{figure}
\includegraphics[width=0.95\columnwidth]{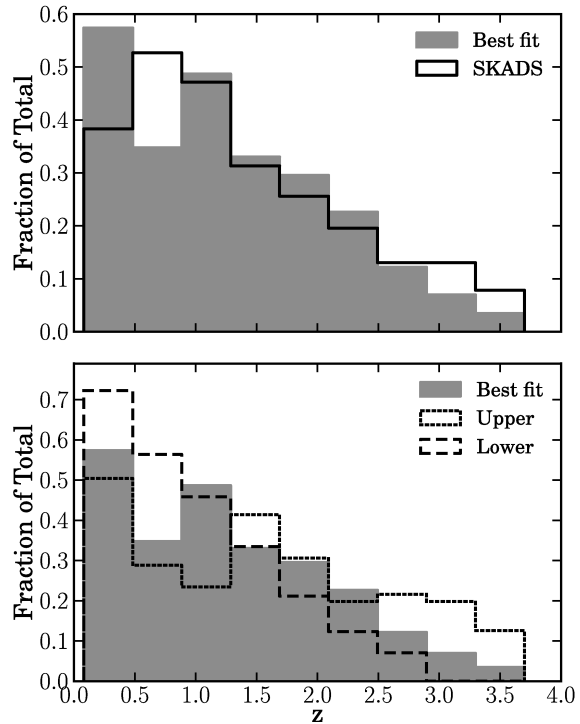}
\caption{
The distribution of the estimated redshifts of the matched radio sources in our sample. The top figure compares the best fit 
redshift  estimates determined by our method to the theoretical redshift distribution in the SKADS simulations \citep{Wilman2008} of a radio survey the same flux 
limits as that described in section \ref{sec:survey}.
The bottom figure compares the best fit distribution to the distributions which would occur if all the sources had redshifts corresponding to 
the upper and lower 68\% confidence intervals of their predicted redshift.
}
\label{fig:dist}
\end{figure}

The highly populated synthetic $K-z$ diagram generated by the Monte Carlo simulations can be used to obtain a photometric redshift probability density function 
for a source of any given K-band magnitude. This is achieved by extracting the redshifts of all the simulated sources in a narrow interval
 about the K-band magnitude in question. A fit to the distribution of these extracted redshifts provides the required probability density function.
These distributions are best-fit by a $\log_{10}$-normal distribution with probability density function given by :
 \begin{equation}
   p(z|K)=\frac{1}{\ln{(10)} z \sqrt{2\pi\sigma^2}}\exp{\{-\frac{[\log_{10}{(z)}-\mu]^2}{2\sigma^{2}}\}}
                                                                                                                   \end{equation}

where $\mu$ and $\sigma$ are the mean and standard deviation for normally distributed random variable $\log_{10}(z)$. The best-fitting 
estimate for $z$ is defined as \begin{equation}
                                z_{est}=10^{\mu}
                               \end{equation}
and the asymmetric 68\% confidence interval is defined as : \begin{equation}
                                                             10^{\mu-\sigma}\leq z_{est} \leq 10^{\mu+\sigma}
                                                            \end{equation}
The distribution of the estimated redshifts for all the radio sources in our survey with $K-$band counterparts is shown in figure \ref{fig:dist}. 
This distribution is compared with the redshift distribution  predicted by the SKADS simulations \citep{Wilman2008,Wilman2010} for a radio continuum survey with flux limits 
comparable to the survey used in this analysis.

This method of redshift estimation does not adjust for the possibility that optical host galaxy luminosity and radio luminosity may be correlated \citep[e.g.][]{Willott2003}.
 This would most likely occur as a result of a correlation between both these properties with the mass of the central black hole. \citet{McLure2004} 
investigated the relationship between radio luminosity and black hole mass and concluded that these properties were only significantly correlated
 for the higher luminosity FRII and HERG sources, and the effect is relatively small. As we are predominantly interested in investigating the evolution of low luminosity sources in this analysis we
are confident that this will not have a significant effect on our conclusions.

Although using this method essentially precludes us from investigating any correlations between galaxy mass and radio emission, \citep[e.g.][]{Willott2003, McLure2004}, it does allow us the possiblility of just using single band photometry to measure the evolution of radio sources, and is therefore much less telescope intensive than using multiband photometry. This is particularly true if one wants to constrain the brighter end of the luminosity function, and therefore driven to use larger survey areas.
 
\section{V/V$_{\rm{max}}$}

The evolution of the comoving space density of sources in a complete sample can be assessed using the non-parametric 
$V/V_{\rm{max}}$ method \citep{Rowan,Schmidt}. Here  $V$ is the volume enclosed by the object and $V_{\rm{max}}$ is the volume enclosed by the object if it were located at 
the redshift $z_{\rm{max}}$ where it's
measured flux drops below the limit of the survey. In order to determine $z_{\rm{max}}$ it is necessary to take into account
 flux density limits of the radio survey and the limiting magnitude of the infrared survey. If $z_{\rm{max}}$ determined by the radio data is larger
than the redshift $z_{opt}$ which would cause the source to drop out of the infrared survey then $z_{\rm{max}}$ = $z_{\rm{opt}}$.  We use our Monte Carlo simulations
of the $K$-z diagram to predict $z_{\rm{opt}}$. We obtain a value of $z_{\rm{opt}}=2.5$ as the mean $z$ corresponding to our limiting K-band magnitude of 19.2.

 If the objects in our sample are uniformly disributed in space then $V/V_{\rm{max}}$ will be uniformly distributed
between 0 and 1 with a mean  value of $\langle V/V_{\rm{max}}\rangle$ = 0.5$\pm(12N)^{-\frac{1}{2}}$, where N is the number of objects in the sample.
 A value of $\langle V/V_{\rm{max}}\rangle >0.5$ implies that  
comoving space density of sources increases at higher redshifts ie negative evolution with cosmic time whereas  $\langle V/V_{\rm{max}}\rangle >0.5$ indicates a decline in space density with increasing 
redshift.

 The original $\langle V/V_{\rm{max}}\rangle$ test was formulated for a single sample complete above a given radio flux density and optical magnitude. However the 
radio survey used in this analysis is 
inhomogeneous in depth \citep{tasse610}, thus it is more appropriate to use the generalised $V_{\rm{e}}/V_{\rm{a}}$ test devised by \citet{Avni} to analyse the evolutionary
properties of this sample.
The generalised test treats the incomplete radio data as a single sample with a variable survey area. This variable survey area depends on the 
flux density of the individual radio source.
 The new test variables are the enclosed volume $V_{\rm{e}}$ and the available volume $V_{\rm{a}}$, the statistical properties of the $V_{\rm{e}}/V_{\rm{a}}$
 variable are identical to those of the $V/V_{\rm max}$ variable in a single flux-limited sample \citep{Avni}.

  To determine the effective survey area as a function of radio flux density we perform a simple simulation. We inject a 1000 radio sources with peak
flux densities ranging from 1-4~mJy into the original map of the radio data. These simulated sources are restricted to the 2.75 degrees$^{2}$ covered by 
the DXS survey. We determine the fraction of these simulated sources which are recovered by the source extraction algorithm SAD in the Astronomical
Image Processing Software (AIPS) as a 
 function of their input flux densities.

\section{V/V$_{max}$ Results}

\begin{figure}
\includegraphics[width=0.95\columnwidth]{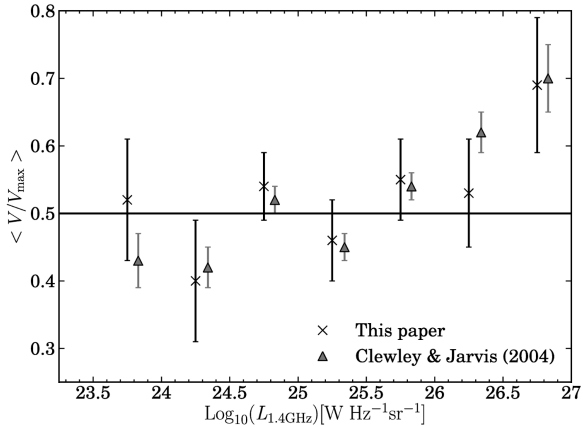}
\caption{ Average $V/V_{\rm{max}}$ in seven different radio luminosity bins compared to the results obtained in \citet{Clewley2004} .}
\label{fig:vvmax}
\end{figure}

In figure \ref{fig:vvmax} we compare the mean $V_{\rm{e}}/V_{\rm{a}}$ statitistics obtained in our analysis with those found in \citet{Clewley2004}. 
We find that our results are in agreement with their analysis which is  based on galaxies selected at 325~Mhz in the Sloan Digital Sky Survey. Our results
also compare very well with the $V/V_{\rm{max}}$  results of \citet{Tassevmax}. Their analysis is based on a combination of 325~MHz and 610~MHz 
radio sources in the XMM-LSS
field, the data used in their analysis overlaps the DXS field studied in this paper so it is encouraging that we obtain a similar result.
The strong evolution of the high luminosity radio sources detected by \citet{Clewley2004} is also in evidence in our figure \ref{fig:vvmax} although 
this result is 
at a lower statistical significance due to the small size of our sample, similarly their is a little evidence that the low power sources are evolving
with cosmic time. The 610~MHz flux densities of our radio sources were converted to 1.4~GHz radio luminosities using the spectral index estimates
 obtained by \citet{tasse610} by comparision with the NVSS survey.       

As discussed in section \ref{reds}, there is some uncertainty in the redshift estimates used in this analysis due to the inherent scatter in the $K-z$ relation. 
To characterise the effect of this source of error on our $V_{\rm{e}}/V_{\rm{a}}$ estimates we use Monte Carlo methods. In these Monte Carlo simulations we assign
each source a redshift estimate determined by the redshift probability density function derived from its' K-band magnitude. We use
 these simulated
redshifts to rederive the $\langle V_{\rm{e}}/V_{\rm{a}} \rangle$ in the seven luminosity bins in figure \ref{fig:vvmax}. We repeat this process 1000 times to obtain the 68\% confidence
interval of the $V_{\rm{e}}/V_{\rm{a}}$ estimates. This confidence interval is indicated in table \ref{tab:vvmax} . The size of this source of error is comparable
to the statistical errors due to the sample size.

\begin{table}
{%

 \begin{center}
\begin{tabular}{l|clll}
\hline
$L_{\rm{1.4GHz}}~[\rm{W~Hz^{-1}sr^{-1}}]$& $\langle V_{\rm{e}}/V_{\rm{a}}\rangle$ & 1$\sigma$ & 1$\sigma$ (MC) & $\langle z \rangle$\\

\hline\hline
23.5-24 & 0.52 & 0.09 & 0.07 & 0.26\\
24-24.5 & 0.40 & 0.09 & 0.06 & 0.47\\
24.5-25 & 0.55 & 0.05 & 0.04 & 0.91\\
25-25.5 & 0.46 & 0.06 & 0.04 & 1.2\\
25.5-26 & 0.55 & 0.06 & 0.05 & 1.7\\
26-26.5 & 0.53 & 0.08 & 0.06 & 1.7\\
26.5-27 & 0.69 & 0.10 & 0.07 & 1.9\\
\hline
 \end{tabular}
 \end{center}
}%
\caption{ Results of $\langle V_{\rm{e}}/V_{\rm{a}}\rangle$ analysis is seven luminosity bins. The column entitled 1$\sigma$ (MC) quotes the uncertainty in our  $\langle V/V_{\rm{max}}\rangle$ estimates
due to the uncertainty in our redshift estimates. The last column gives the mean $z$ of sources in each bin. }
\label{tab:vvmax} 
\end{table}

\begin{figure}
\includegraphics[width=0.85\columnwidth]{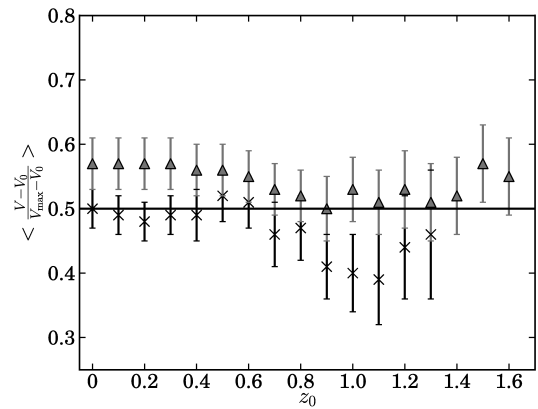}
\caption{
The banded $V/V_{\rm{max}}$ in two luminosity bins. The bin with $L_{\rm{1.4GHz}}>10^{25}~\rm{W~Hz^{-1}sr^{-1}}$ is represented by grey triangles and the black crosses represent
 the  $L_{\rm{1.4GHz}}<10^{25}~\rm{W~Hz^{-1}sr^{-1}}$ bin. }

\label{fig:band}
\end{figure}

 To further our investigation of this evolutionary behaviour we determine the banded $V_{\rm{e}}/V_{\rm{a}}$ statistic in two luminosity bins. A high 
luminosity bin corresponding to sources with $L_{\rm{1.4GHz}}>10^{25}~\rm{W~Hz^{-1}sr^{-1}}$ and a low luminosity bin with  $L_{\rm{1.4GHz}}<10^{25}~\rm{W~Hz^{-1}sr^{-1}}$.
 The banded test calculates the mean $V_{\rm{e}}/V_{\rm{a}}$ for all sources with $z>z_{0}$ for a range of $z_{0}$ values. All evolution below $z_{0}$ is
thus masked out of the analysis allowing us to determine whether the evolution of these two populations of radio sources changes as a function of redshift. The
results of the banded test are presented in figure \ref{fig:band}. The error in the banded test as a result of the uncertainty in our redshift estimates
is also calculated using Monte Carlo methods and presented in table \ref{tab:vmaxband}. 
  
The results in figure \ref{fig:band} indicate that there is  little evidence of evolution in the low power radio sources  out
to a redshift of $z\sim0.8$. Beyond this redshift there is a slight decline in the average $V_{\rm{e}}/V_{\rm{a}}$  of this population however the errors
derived from our Monte Carlo estimates increase significantly at these higher redshifts. The stronger evolution of the high luminosity sources
can also be seen in this figure. The mean $V_{\rm{e}}/V_{\rm{a}}$ for the low and high luminosity bins is calculated as $0.5\pm 0.03$ and $0.57\pm 0.04$ respectively.%

\begin{table}

{%
\begin{center}
\begin{tabular}{l|cll|cll}
\hline
× & \mc{3}{c}{$L_{\rm{1.4GHz}}<10^{25}~\rm{W~Hz^{-1}sr^{-1}}$} & \mc{3}{c}{$L_{\rm{1.4GHz}}>10^{25}~\rm{W~Hz^{-1}sr^{-1}}$}\\
$z_{0}$ & $\langle V_{\rm{e}}/V_{\rm{a}}\rangle$ & 1$\sigma$ & 1$\sigma$(MC) &  $\langle V_{\rm{e}}/V_{\rm{a}} \rangle$  & 1$\sigma$ & 1$\sigma$ (MC)\\
\hline\hline
0 & 0.50 & 0.03 & 0.02 & 0.57 & 0.04 & 0.02\\
0.2 & 0.48 & 0.03 & 0.02 & 0.57& 0.04 & 0.02\\
0.4 & 0.49 & 0.04 & 0.04 & 0.56 & 0.04 & 0.03\\
0.6 & 0.51 & 0.04 & 0.05 & 0.55 & 0.04 & 0.03\\
0.8 & 0.47 & 0.05 & 0.07 & 0.52 & 0.04 & 0.03\\
1.0 & 0.40 & 0.05 & 0.14 & 0.53 & 0.05 & 0.04\\
1.2 & 0.44 & 0.08 & 0.24 & 0.53 & 0.05 & 0.05\\
1.4& × & × &  & 0.52 & 0.06 & 0.05\\
1.6 & × & × & × & 0.55& 0.05 & 0.06\\
\hline
\label{tab:vmaxband}
\end{tabular}
\end{center}
}%
\caption{Results of the banded $\langle V_{\rm{e}}/V_{\rm{a}}\rangle$ test. The columns entitled 1$\sigma$ (MC) list the Monte Carlo
estimates of the error in our $\langle V_{\rm{e}}/V_{\rm{a}}\rangle$ estimates due to scatter in the $K-z$ relation. }

\end{table}

\section{Radio Luminosity Function}
Another standard means of quantifying the cosmic evolution of radio sources  is to compare the measured radio luminosity function (LF) of 
these sources at different cosmic epochs. Changes in the luminosity function $\Phi_{z}(L)$ are usually modelled as being due to one of two simplified evolution
scenarios, either pure density evolution or pure luminosity evolution. In pure density evolution the luminosity distribution of the radio sources
is considered fixed while the number density of radio sources changes with redshift. In this scenario the measured LF at a given redshift  
 $\Phi_{z}(L)$ is related to the local
luminosity function $\Phi_{0}(L)$ via the following equation :  \begin{equation} \Phi_z(L)=(1+z)^{\alpha_{D}}\Phi_{0}(L) \end{equation}
In pure luminosity evolution the radio sources undergo a change in their luminosity with time, this evolution is parameterized as:
\begin{equation} \Phi_{z}(L)=\Phi_{0}\left(\frac{L}{(1+z)^{\alpha_{L}}}\right)\end{equation}

\begin{figure}
 \includegraphics[width=0.99\columnwidth]{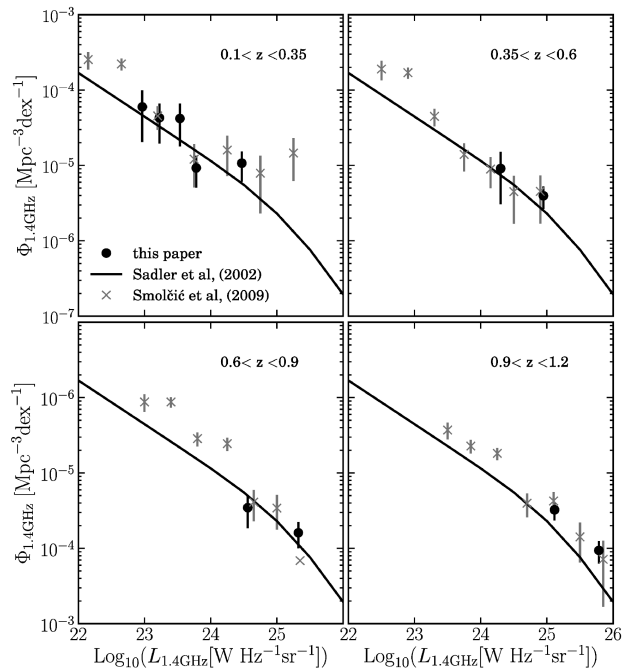}
\caption{The measured radio luminosity function in four redshift bins compared to the results of \citet{Smol2009}. The solid line represents the analytic form of the local
luminosity function derived in \citet{Sadler2002}.}
\label{fig:rlf}
\end{figure}

We calculate the $\Phi_{z}(L)$ of our sample in four redshift bins using the standard 1/V$_{max}$ method of \citet{Schmidt}. Thus the radio LF and its corresponding error is calculated for a particular luminosity bin as:
  \begin{align}
  & \Phi(L)=\sum_{j=1}^{N}\frac{1}{V_{\rm{max}}^{j}}   &  \sigma  &= \sqrt{\sum_{j=1}^{N}\left(\frac{1}{V_{\rm{max}}^{j}}\right)^{2}}
                                                                   \end{align}
where the j index runs over all sources in the luminosity bin L-$\Delta L\rightarrow$L+$\Delta L$. We adjust the V$_{max}$ values to take into account
 the redshift limits of the four redshift bins for which the luminosity functions were calculated. The results of our analysis are presented in
 figure \ref{fig:rlf} and compared
with the  LF of radio sources in the VLA-COSMOS survey derived by \citet{Smol2009}. Our luminosity function estimates appear to be in good agreement with the 
results of this previous study.

\citet{Smol2009} characterise the evolution of the radio sources in the VLA-COSMOS survey by adopting the LF given by \citet{Sadler2002} as the best
 representative of their measured local
luminosity function and using a least squares fit to derive the $\alpha_{D}$ and $\alpha_{L}$ parameters for pure density and pure luminosity evolution of this LF.
The analytic form of the radio LF given in \citet{Sadler2002} was obtained from radio galaxies in the NVSS survey matched with optical
counterparts in the 2dF Galaxy Redshift Survey. 

 The fitting procedure of \citet{Smol2009} yeilded estimates of
 $\alpha_{D}=1.1\pm0.1$ and $\alpha_{L}=0.8\pm0.1$ where the errors were derived from the distribution of $\chi^{2}$ in the least squares fit. 
Employing this same method we obtain values for the density and luminosity evolution parameters of $\alpha_{D}=0.6\pm0.1$ and 
$\alpha_{L}=0.8\pm0.2$. These estimates are fairly consistent with the estimates of \citet{Smol2009} and imply an increase in the
 co-moving space density of low luminosity radio sources by a factor of $\sim1.5$ in the interval between z=$0-0.8$. 

The consistency of our results
with those of \citet{Smol2009} provides encouranging evidence that the $K-z$ relation can be used to investigate the radio luminosity function. This 
could provide a useful alternative technique for studying the LF in the absence of multi-band photometric redshift estimates. 

\section{Conclusion} 
We have used the non-parametric $V/V_{\rm{max}}$ test and the radio luminosity function to investigate the the cosmic evolution of low power radio sources in
the XMM-LSS field. Previous investigations of the evolution of low-power radio sources have found evidence of mild evolution taking place in the
redshift range $z=0-1.3$. \citet{Donoso} find that the co-moving density of low luminosity sources ($L_{\rm{1.4GHz}}<10^{25}~\rm{W~Hz^{-1}sr^{-1}}$) in the MegaZ-Luminous Red Galaxy catalogue
 increases by a 
factor of 1.5
between $z=0.1-0.55$. Similarly \citet{Sadler} find evidence that low luminosity sources experience  mild evolution with  
 an increase in their number density by a factor of $\sim$2 at z=0.55. This is significantly less than 
the strong evolution detected in the high power
radio sources whose number densities are enhanced by a factor of 10 in the same cosmic timeframe. The results of \citet{Sadler} seem to rule out the
no-evolultion scenario. \citet{Gendre2010} find evidence that higher luminosity, $L_{\rm{1.4GHz}} \geq 10^{24.5}~\rm{W~Hz^{-1}sr^{-1}}$ FRI

  \citet{Smol2009} find mild evolution of the low power AGN in the VLA-COSMOS survey out to $z\sim1.3$. Our results are broadly consistent with
these previous works and imply density enhancements by a factor $\sim1.5$ at z=0.8. This is slightly less than the evolution 
implied in \citet{Sadler}, but in fairly good agreement with the estimates in \citet{Smol2009} and \citet{Donoso}. We also find evidence that the low power radio AGN are evolving more slowly than their high power counterparts in the redshift range $z=0-0.8$ and tentative evidence that this
separation in evolutionary behaviour persists to $z=1.2$. 


The division in the FRI/FRII classification system  and  the LERG/HERG system occurs at roughly the same luminosity threshold of
 $L_{\rm{1.4GHz}}\sim10^{25}$.
As our analysis categorises our radio sources solely on their  luminosity we are unable to determine whether the separation  in evolutionary behaviour is due to the FRI/FRII dichotomy or to differences in black hole accretion
modes in HERGS and LERGS. 

Our results demonstrate that using the $K-$band (or similar wavelength), combined with radio surveys,
 is a viable route to investigating the evolution of the radio source population, at least up to $z \sim 1.2$. 
In the future, all-sky radio surveys such as those carried out with the Low-Frequency Array \citep[LOFAR; ][]{Morganti2010} and 
the Australian Square Kilometre Array Pathfinder (ASKAP) telescope, combined with the UKIDSS large area survey and the VISTA Hemisphere 
Survey, as well as Wide-field Infrared Survey Explorer (WISE), will enable us to pin down the evolution of the radio source population to a much higher degree of accuracy.

\section*{ACKNOWLEDGEMENTS} 
Kim McAlpine gratefully acknowledges financial
support received from the South African SKA project and the National Research Foundation.  
MJJ ackowledges the support of a RCUK fellowship, and visitor funding from the  University of Cape Town and the South African SKA project.
This work is based in part on data obtained as part of the UKIRT Infrared Deep Sky Survey.



\end{document}